\documentclass{article}%
\usepackage{graphicx}
\usepackage{amsmath}%
\usepackage{amsfonts}%
\usepackage{amssymb}
\newtheorem{theorem}{Theorem}
\newtheorem{acknowledgement}[theorem]{Acknowledgement}

\begin{document}

\title{Quantum Time Arrows, Semigroups and Time-Reversal in Scattering}
\author{Robert C. Bishop$^{a,b}$\\$^{a}$Abteilung f\"{u}r Theorie und Datenanalyse, Institut\\f\"{u}r Grenzgebiete der Psychologie, Wilhelmstrasse\\3a, D-79098 Freiburg, Germany\\$^{b}$Permanent Address: Department of Philosophy,\\Logic and Scientific Method, The London School\\of Economics, Houghton St., London,\\WC2A 2AE, United Kingdom\\Accepted for publication in the \textit{International Journal of Theoretical Physics}}
\date{}
\maketitle

\begin{abstract}
Two approaches toward the arrow of time for scattering processes have been
proposed in rigged Hilbert space quantum mechanics. One, due to Arno Bohm,
involves preparations and registrations in laboratory operations and results
in two semigroups oriented in the forward direction of time. The other,
employed by the Brussels-Austin group, is more general, involving excitations
and de-excitations of systems, and apparently results in two semigroups
oriented in opposite directions of time. It turns out that these two time
arrows can be related to each other via Wigner's extensions of the spacetime
symmetry group. Furthermore, their are subtle differences in causality as well
as the possibilities for the existence and creation of time-reversed states
depending on which time arrow is chosen.

\begin{acknowledgement}
I would like to thank I. Antoniou, H. Atmanspacher, A. Bohm, R. De La Madrid
and S. Wickramasekara for illuminating discussions. Any remaining confusions
are my own.

\end{acknowledgement}
\end{abstract}

\section{\noindent Introduction}

In the standard formulation of nonrelativistic quantum mechanics, the time
evolution of systems is governed by a one-parameter group of unitary
operators
\begin{equation}
U(t)=e^{-iHt}%
\end{equation}

\noindent on a Hilbert space (HS) [18], where $H$ represents the Hamiltonian
and Planck's constant has been set to one. Any evolution governed by (1) is
time-reversal invariant\footnote{Time-reversal invariance means that if
$\phi(t)$ is a solution of the quantum mechanical equations of motion, then so
is $\phi(-t)$.} and irreversibility\footnote{A process is \textit{reversible}
if the temporal succession of its states $\phi_{1}$, $\phi_{2}$,..., $\phi
_{n}$ can occur as well as the reverse sequence of states $\phi_{n}^{T}$,
$\phi_{n-1}^{T}$,..., $\phi_{1}^{T}$, where $T$ is a time-reversal operation;
otherwise it is \textit{irreversible}.} usually enters in due to an extrinsic
act of measurement or other interaction with an environment [19, 21]. This
approach, however, has some undesirable features: (1) the observed exponential
decay in various quantum experiments is considered as being only approximately
exponential; (2) no appropriate eigenvectors describing decaying states (e.g.
Gamow vectors and Dirac states) are elements of HS; (3) there is a tendency to
treat metastable states such as resonances or decaying states as transients
rather than as states of autonomous microphysical systems; (4) no intrinsic
forms of irreversibility--where irreversible behavior originates in the
dynamics of a physical system without explicit reference to an environment
[1]--can be appropriately modeled nor can appropriate initial conditions for
such irreversible processes be formulated rigorously in HS.

For these, among other reasons [3, 4, 8], theories of rigged Hilbert space
(RHS) quantum mechanics--a generalization of the HS version--were developed
[2, 4, 5]. A RHS, or Gel'fand triplet [12, 11], \ is the triple of spaces%
\begin{equation}
\Phi\subset\mathcal{H}\subset\Phi^{\times}\text{ ,}%
\end{equation}

\noindent where $\mathcal{H}$ is a HS with the standard norm topology,
$\tau_{\mathcal{H}}$, $\Phi$ is a vector space with a topology, $\tau_{\Phi}$,
stronger than $\tau_{\mathcal{H}}$ and $\Phi^{\times}$ is the dual space of
continuous linear functionals on $\Phi$. A RHS provides an appropriate setting
for studying intrinsically irreversible processes because it naturally
accommodates semigroup evolutions and the initial and boundary conditions
appropriate to such evolutions [8].

In the context of scattering theory, two arrows of time intrinsic to the
dynamics of quantum systems have been proposed within RHS quantum mechanics.
One, due to Bohm [7,8], involves preparations and registrations in laboratory
operations, resulting in semigroups oriented in the forward direction of time.
The other, originally proposed by George [13] and employed by the
Brussels-Austin group, is more general involving excitations and
de-excitations of systems, resulting in semigroups apparently oriented in
opposite directions of time. I will briefly review these two quantum arrows of
time and then examine their relationship under time-reversal.

\section{States and Observables}

A typical scattering experiment consists of an accelerator, which prepares a
projectile in a particular state, a target and detectors. The total
Hamiltonian modeling the interaction of the particle with the target is,
therefore, $H$ = $H_{o}$ + $V$, where $H_{o}$ represents the free particle
Hamiltonian and $V$ the potential in the interaction region. The vectors
representing growing and decaying states are associated with the resonance
poles of the analytically continued S-matrix [14].

The \textit{preparation/registration} arrow of time [8] is fundamental to
Bohm's analysis of resonance states. The key intuition behind this arrow is
that no observable properties of a state can be measured unless the state has
first been prepared. Following Ludwig [16, 17, 8], an in-state of a particular
quantum system (considered as an ensemble of individual systems such as
elementary particles) is prepared by a preparation apparatus (considered
macrophysical). The detector (considered classical) registers so-called
out-states of post-interaction particles. In-states are taken to be elements
$\phi\in\Phi_{-}$ and observables are taken to be elements $\psi\in\Phi_{+}$.
(Decaying states, such as the Dirac, Lippman, Schwinger kets and Gamow
vectors, are elements of $\Phi_{\pm}^{\times}$). This leads to a distinction
between prepared states and observables, each described by a separate RHS [8]:%
\begin{subequations}
\begin{align}
\Phi_{-}  &  \subset\mathcal{H}\subset\Phi_{-}^{\times}\text{ }\\
\Phi_{+}  &  \subset\mathcal{H}\subset\Phi_{+}^{\times}\text{ ,}%
\end{align}

\noindent where $\Phi_{-}$ is the Hardy space of the lower complex energy
half-plane intersected with the Schwartz class functions and $\Phi_{+}$ is the
Hardy space of the upper complex energy half-plane intersected with the
Schwartz class functions [8]. As Bohm and Gadella [7] demonstrate, some
elements of the generalized eigenstates in $\Phi_{-}^{\times}$ and $\Phi
_{+}^{\times}$ correspond to exponentially growing and decaying states
respectively. The semigroups governing these states are\footnote{If $U(t)$ is
a unitary operator on $\mathcal{H}$ and $\Phi\subset\mathcal{H}\subset
\Phi^{\times}$, then $U^{\dagger}$ can be \textit{extended} to $\Phi^{\times}%
$\ provided that (1) $U$ leaves $\Phi$ invariant, i.e. $U$: $\Phi
\rightarrow\Phi$, and (2) $U$ is continuous on $\Phi$ with respect to the
topology $\tau_{\Phi}$. The operator $U^{\times}$\ denotes the
\textit{extension} of the HS operator $U^{\dagger}$ to $\Phi^{\times}$\ and is
defined by $\langle U\phi|F\rangle=\langle\phi|U^{\times}F\rangle$ for all
$\phi\in\Phi$ and $F\in\Phi^{\times}$. When the group operator $U^{\dagger}$
is extended to $\Phi^{\times}$, continuity requirements force the operators
$U^{\times}$ to be semigroups defined only on the temporal half-domains [7].}%
\end{subequations}
\begin{subequations}
\begin{align}
\langle\phi|U^{\times}|Z_{R}^{\ast}\rangle &  =e^{-iE_{R}t}e^{\frac{\Gamma}%
{2}t}\langle\phi|Z_{R}^{\ast}\rangle\text{ }t\leq0\text{, }t:-\infty
\rightarrow0\\
\langle\psi|U^{\times}|Z_{R}\rangle &  =e^{-iE_{R}t}e^{-\frac{\Gamma}{2}%
t}\langle\psi|Z_{R}\rangle\text{ }t\geq0\text{, }t:0\rightarrow\infty\text{,}%
\end{align}
where $E_{R}$ represents the total resonance energy, $\Gamma$ represents the
resonance width, $Z_{R}$ represents the pole at $E_{R}-i\frac{\Gamma}{2}$,
$Z_{R}^{\ast}$ represents the pole at $E_{R}+i\frac{\Gamma}{2}$, $|Z_{R}%
^{\ast}\rangle\in\Phi_{-}^{\times}$ represents a growing Gamow vector and
$|Z_{R}\rangle\in\Phi_{+}^{\times}$ represents a decaying Gamow vector. The
$t<0$ semigroup is identified as future-directed along with $|Z_{R}^{\ast
}\rangle$ as a forming/growing state. The $t>0$ semigroup is identified as
future-directed along with $|Z_{R}\rangle$ as a decaying state\footnote{Note
that the eigenvectors plus the semigroup property are insufficient to
determine the temporal direction of evolution. These identifications involve
further physical justification.}.

In their discussion of scattering and resonance phenomena, Antoniou and
Prigogine also apply the RHS framework, using the Hardy class functions as a
natural function space for their analysis [2]. Antoniou and Prigogine adopt a
time arrow somewhat different from Bohm [3]: excitations are interpreted as
events taking place before $t=0$ while de-excitations are interpreted as
events taking place after $t=0$. This time arrow leads to a natural splitting
of the RHS: excitations (e.g. formation of unstable states) are considered as
past-oriented and are associated with $\phi_{+}\in$ $\Phi_{+}^{\times}$ in the
upper half-plane, while de-excitations (e.g. decay of ustable states) are
considered as future-oriented and are associated with $\phi_{-}\in\Phi
_{-}^{\times}$ in the lower half-plane.\footnote{Note that the roles of the
upper and lower Hardy class function spaces is reversed with respect to Bohm's
approach. This has only mathematical import. The differences in phase factors
between (4) and (5) are due to the fact that in the former, states evolve in
the Schr\"{o}dinger picture while observables evolve in the Heisenberg
picture, while in the latter, only the Schr\"{o}dinger picture is used.} The
semigroups governing decaying states as identified by the Brussels-Austin
group are
\end{subequations}
\begin{subequations}
\begin{align}
\langle\phi_{+}|U^{\times}|Z_{R}^{\ast}\rangle &  =e^{iE_{R}t}e^{\frac{\Gamma
}{2}t}\langle\phi_{+}|Z_{R}^{\ast}\rangle\text{ }t<0\text{, }t:-\infty
\leftarrow0\\
\langle\phi_{-}|U^{\times}|Z_{R}\rangle &  =e^{-iE_{R}t}e^{-\frac{\Gamma}{2}%
t}\langle\phi_{-}|Z_{R}\rangle\text{ }t>0\text{, }t:0\rightarrow\infty\text{.}%
\end{align}
The Brussels-Austin Group identifies the $t<0$ semigroup as evolving states
into the past along with $|Z_{R}^{\ast}\rangle$ as decaying states, and the
$t>0$ semigroup as evolving states into the future along with $|Z_{R}\rangle$
as decaying states.

\section{Time-reversal}

Following Wigner [20], the time-reversal operator, $R(t)$, is the HS
representation of the physical spacetime transformation%
\end{subequations}
\begin{equation}
R:(\vec{x},t)\rightarrow(\vec{x},-t)\text{.}%
\end{equation}

\noindent Therefore, $R$ is an element of a co-representation of the extended
Galilei symmetry group [10] for nonrelativistic spacetime (extended
Poincar\'{e} group for relativistic spacetime). These representations must be
unitary and linear except for $R$, which is antilinear. With these properties,
$R$ fulfils%
\begin{subequations}
\begin{align}
RP_{i}R^{-1}  &  =-P_{i}\\
RJ_{i}R^{-1}  &  =-J_{i}\\
RK_{i}R^{-1}  &  =K_{i}\\
RHR^{-1}  &  =H\\
RSR^{-1}  &  =S^{\dagger}=S^{-1}\text{,}%
\end{align}

\noindent where $P_{i}$, $J_{i}$, $K_{i}$, $H$ and $S$ are the momentum,
angular momentum, Lorentz boost, energy and S-matrix operators respectively
[9]. The relation (7e) is experimentally tested in the form of the reciprocity
relation, but it should be pointed out that (7) is formulated in terms of
observables, not states.

However, there is one more technicality to discuss before examining the
application of $R$ to the states and observables of \S2. Wigner originally
derived the properties of $R$ for the spacetime symmetry group extended by
time inversions and studied the parity inversion operator $\Sigma$ and the
total inversion operator $T$ in combination with $R$ [20]. The parity
inversion operator is unitary so its phase can be chosen such that $\Sigma
^{2}$ $=I$ (the identity operator), while $T$ and $R$ are both anti-unitary,
so that the associative law for group multiplication dictates that
$R^{2}=\varepsilon_{R}I$ and $T^{2}=\varepsilon_{T}I$, where $\varepsilon
_{R}=\pm1$ and $\varepsilon_{T}=\pm1$. The phase of $T$ can be chosen so that
$T=\Sigma R$ (where the order of application of $\Sigma$ and $R$ is physically
immaterial). The extension of the spacetime symmetry group is summarized in
Table 1.\medskip

$\underset{\text{Table 1. Properties of the spacetime symmetry group.}}{%
\begin{tabular}
[c]{|c|c|c|c|c|}\hline
$\varepsilon_{R}$ & $\varepsilon_{T}$ & $\Sigma$ & $R$ & $T$\\\hline
$(-1)^{2j}$ & $(-1)^{2j}$ & $1$ & $C$ & $C$\\\hline
$-(-1)^{2j}$ & $(-1)^{2j}$ & $\left(
\begin{array}
[c]{cc}%
1 & 0\\
0 & -1
\end{array}
\right)  $ & $\left(
\begin{array}
[c]{cc}%
0 & C\\
-C & 0
\end{array}
\right)  $ & $\left(
\begin{array}
[c]{cc}%
0 & C\\
C & 0
\end{array}
\right)  $\\\hline
$(-1)^{2j}$ & $-(-1)^{2j}$ & $\left(
\begin{array}
[c]{cc}%
1 & 0\\
0 & -1
\end{array}
\right)  $ & $\left(
\begin{array}
[c]{cc}%
0 & C\\
C & 0
\end{array}
\right)  $ & $\left(
\begin{array}
[c]{cc}%
0 & C\\
-C & 0
\end{array}
\right)  $\\\hline
$-(-1)^{2j}$ & $-(-1)^{2j}$ & $\left(
\begin{array}
[c]{cc}%
1 & 0\\
0 & 1
\end{array}
\right)  $ & $\left(
\begin{array}
[c]{cc}%
0 & C\\
-C & 0
\end{array}
\right)  $ & $\left(
\begin{array}
[c]{cc}%
0 & C\\
-C & 0
\end{array}
\right)  $\\\hline
\end{tabular}
\ }\medskip$

\noindent The index $j$ refers to the spin of the particle being considered
while $C$ is an operator whose $(2j+1)$-dimensional matrix has the elements
$c_{%
\mu
,\nu}=(-1)^{j+%
\mu
}\delta_{%
\mu
,\nu}$, where $-j\leq%
\mu
$ and $\nu\leq j$. In the first representation, where $\varepsilon
_{R}=\varepsilon_{T}=(-1)^{2j}$, there are no changes to the underlying vector
space. This is the typical case discussed in quantum mechanics (and
relativistic quantum field theory). The other three representations, however,
exhibit a doubling of the vector spaces. In order to track this space
doubling, let the index $r=0,1$ label the rows and columns of the operator
matrices in Table 1.

\section{Time-reversed States and Observables}

Although no quantum fields have been constructed for representations two and
three of Table 1 (indeed they are highly problematic), Bohm and co-workers
have constructed models for the fourth representation by applying R to the
states and observables in (4) [6, 9]. First, consider the growing Gamow
vectors for, $\phi^{r=0,\times}\in\Phi_{-}^{r=0,\times}$. Applying $R$ yields%
\end{subequations}
\begin{equation}
R\phi^{r=0,\times}=\psi^{r=1,\times}\in\Phi_{+}^{r=1,\times}\text{.}%
\end{equation}
Similarly for the decaying Gamow vectors, $\psi^{r=0,\times}\in\Phi
_{+}^{r=0,\times}$, applying $R$ yields%
\begin{equation}
R\psi^{r=0,\times}=\phi^{r=1,\times}\in\Phi_{-}^{r=1,\times}\text{.}%
\end{equation}
The transformation properties of $R$ may be summarized as $R:\Phi_{\pm
}^{r=0,\times}\rightarrow\Phi_{\mp}^{r=1,\times}$. The temporal evolution of
these time-reversed vectors is also given by semigroups. Identify $r=0$ with
the scattering experiment as normally carried out in the laboratory and $r=1$
with the time-reversed situation. Then $U^{\times}(t)\langle\phi
,r=0|Z_{R}^{\ast},r=0\rangle\in\Phi_{-}^{r=0,\times}$, a growing Gamow vector
representing a preparable state for $t\leq0$, is transformed under $R$ into
$U^{\times}(-t)\langle\psi,r=1|Z_{R},r=1\rangle\in\Phi_{+}^{r=1,\times}$,
where%
\begin{equation}
e^{iE_{R}t}e^{-\frac{\Gamma}{2}t}\langle\psi,r=1|Z_{R},r=1\rangle
\end{equation}
is restricted to the time domain $t\geq0$ by continuity requirements. In the
case of $|Z_{R}^{\ast},r=0\rangle$, time runs from $-\infty$ to $0$; in
contrast, for $|Z_{R},r=1\rangle$, time runs from $\infty$ to $0$, meaning
that it represents a Gamow vector that increases as $t$ decreases. Similarly,
$U^{\times}(t)\langle\psi,r=0|Z_{R},r=0\rangle\in\Phi_{+}^{r=0,\times}$, a
decaying Gamow vector representing observables for $t\geq0$, is transformed
under $R$ into $U^{\times}(-t)\langle\phi,r=1|Z_{R}^{\ast},r=1\rangle\in
\Phi_{-}^{r=1,\times}$, where%
\begin{equation}
e^{iE_{R}t}e^{\frac{\Gamma}{2}t}\langle\phi,r=1|Z_{R}^{\ast},r=1\rangle
\end{equation}
is restricted to the time domain $t\leq0$ by continuity requirements. In the
case of $|Z_{R},r=0\rangle$, time runs from $0$ to $\infty$; in contrast, for
$|Z_{R}^{\ast},r=1\rangle$, time runs from $0$ to $-\infty$, meaning that it
represents a Gamow vector that decays as -$t$ increases. These results are
summarized in Table 2.

\medskip

$\underset{\text{Table 2. Properties of the Bohm/Gadella Gamow vectors under
}R(t)\text{.}}{%
\begin{tabular}
[c]{|l|l|l|}\hline
Growing & $\langle\phi,r=0|Z_{R}^{\ast},r=0\rangle$ & $\langle\psi
,r=1|Z_{R},r=1\rangle$\\\hline
Vectors & $t\leq0$, $t:$ $-\infty\rightarrow0$ & $t\geq0$, $t:0\leftarrow
\infty$\\\hline
&  & \\\hline
Decaying & $\langle\psi,r=0|Z_{R},r=0\rangle$ & $\langle\phi,r=1|Z_{R}^{\ast
},r=1\rangle$\\\hline
Vectors & $t\geq0$, $t:$ $0\rightarrow\infty$ & $t\leq0$, $t:-\infty
\leftarrow0$\\\hline
\end{tabular}
\ \ \ \ }$

\medskip

The time-reversed situation in the Brussels-Austin approach have not been
discussed in the literature. Using the transformation rules as appropriate,
the temporal evolution of the time-reversed vectors can be determined.
However, notice that the eigenvectors in (5) are identified with decaying
states. It can be easily seen that (5b) is the time-reversal of (5a) under
$R$, but the label $r$ associated with vector space doubling remains to be
identified. If we assume that the preparation/registration arrow is a special
case of the excitation/de-excitation arrow--that is, that laboratory
preparations are particular types of excitations and the detections of
decaying states are particular types of de-excitations [3]--then (5a) can be
identified with the $r=1$ and (5b) with the $r=0$ regimes respectively
(compare with (4b)).

What remains is to examine the eigenvectors representing growing states in the
Brussels-Austin approach. To each de-excitation in (5) there is a
corresponding excitation represented by an eigenvector in the opposite
temporal half-plane. For the $r=0$ regime, a growing eigenvector of the form
\begin{equation}
e^{iE_{R}t}e^{\frac{\Gamma}{2}t}\langle\phi_{+},r=0|Z_{R}^{\ast}%
,r=0\rangle\text{,}%
\end{equation}
corresponds to eigenstate (5b), where (12) is restricted to the time domain
$t<0$ by continuity requirements. This state is represented by a Gamow vector
that grows as $-t$ decreases. Similarly, for the $r=1$ regime, a growing
eigenstate of the form%
\begin{equation}
e^{-iE_{R}t}e^{-\frac{\Gamma}{2}t}\langle\phi_{-},r=1|Z_{R},r=1\rangle\text{,}%
\end{equation}
corresponds to eigenvector (5a), where (13) is restricted to the time domain
$t>0$ by continuity requirements. This state is represented by a Gamow vector
that grows as $t$ decreases. These results are summarized in Table 3.

\medskip

$\underset{\text{Table 3. Properties of the Brussels-Austin Gamow vectors
under }R(t)\text{.}}{%
\begin{tabular}
[c]{|l|l|l|}\hline
Growing & $\langle\phi_{+},r=0|Z_{R}^{\ast},r=0\rangle$ & $\langle\phi
_{-},r=1|Z_{R},r=1\rangle$\\\hline
Vectors & $t<0$, $t:-\infty\rightarrow0$ & $t>0$, $t:0\leftarrow\infty
$\\\hline
&  & \\\hline
Decaying & $\langle\phi_{-},r=0|Z_{R},r=0\rangle$ & $\langle\phi_{+}%
,r=1|Z_{R}^{\ast},r=1\rangle$\\\hline
Vectors & $t>0$, $t:0\rightarrow\infty$ & $t<0$, $t:-\infty\leftarrow
0$\\\hline
\end{tabular}
\ \ \ \ \ \ \ }\medskip$

The Bohm and Brussels-Austin groups appear to be working with the same
eigenvectors and semigroups in their analyses of scattering. (5a) and (5b) are
time-reversed images of each other and, when paired with their corresponding
growing vectors, are easily related to those of Bohm and co-workers (compare
Tables 2 and 3), which is not immediately apparent when comparing (4) and (5)
without taking time reversal and vector space doubling into account.

\section{The Possibility of Time-reversed States}

It has been suggested that (5a) be disregarded because it is inconsistent with
observations or because of other consistency requirements such as the need for
devices to communicate [2, 3]. Does the consideration of time-reversed states
in the light of vector space doubling lead to new arguments for disregarding (5a)?

\subsection{Physical Considerations}

Lee [15] discusses the following problem with time-reversed quantum states.
Consider a $\bar{\mu}$-meson at rest with its spin $\mathbf{s}_{\mu}$ in the
up direction. It decays as%
\begin{equation}
\bar{\mu}\rightarrow e^{-}(L)+\bar{\nu}_{e}(R)+\nu_{\mu}(L)\text{,}%
\end{equation}
where the electron, electron anti-neutrino and $\mu$ neutrino are emitted with
helicities $-1/2$, $1/2$ and $-1/2$ respectively, denoted by the letters $L$
and $R$ indicating the helicities. Neglecting the electronic mass and assuming
that the final momenta of $e^{-}$, $\bar{\nu}_{e}$ and $\nu_{\mu}$ are
$\mathbf{P}_{e}$, $\mathbf{P}_{\bar{\nu}}$ and $\mathbf{P}_{\nu}$,
respectively, the time-reversed process would be%
\begin{equation}
e^{-}(L)+\bar{\nu}_{e}(R)+\nu_{\mu}(L)\rightarrow\bar{\mu}\text{,}%
\end{equation}
where the initial states of $e^{-}$, $\bar{\nu}_{e}$ and $\nu_{\mu}$ have
momenta $-\mathbf{P}_{e}$, $-\mathbf{P}_{\bar{\nu}}$ and $-\mathbf{P}_{\nu}$
respectively. If time reversal holds, then (15) should lead to a final state
with $\bar{\mu}$ at rest. Also (15) should produce a final spin $\mathbf{s}%
_{\mu}^{^{\prime}}=-\mathbf{s}_{\mu}$, but this is not generally the case in
quantum mechanics. For example, if the momenta of $\bar{\nu}_{e}$ and
$\nu_{\mu}$ are parallel in (14), then conservation of total angular momentum
in (15) requires that $\mathbf{s}_{\mu}^{^{\prime}}$ lie in the same direction
as the initial electron spin, which is typically different from that of
$-\mathbf{s}_{\mu}$. In the more general case, where the directions of the
momenta in (14) are arbitrary, the final spin $\mathbf{s}_{\mu}^{^{\prime}%
}=-\mathbf{s}_{\mu}$ in (15) is only possible if the momentum and spin of all
three leptons are simultaneously reversed in all possible directions while
maintaining the appropriate phase relations among their wave amplitudes. The
latter would require the creation of three perfectly coherent incoming
spherical waves in the midst of the many degrees of freedom involved.

Producing such a state in laboratory situations (preparation/registration
arrow) is clearly impossible because the precision required to produce such
coherent incoming spherical waves, as well as the control over the environment
it entails, exceeds our engineering capabilities (presuming we knew how to
produce such phase-related time-reversed waves). For more general unstable
quantum processes (excitation/de-excitation arrow), it is not clear that
time-reversed growing states associated with the $r=1$ regime can be ruled out
so easily. Though highly improbable, perhaps some kinds of singular events can
produce the kinds of time-reversed processes meeting such stringent requirements.

There is a related question as to why we live in a universe where the
overwhelming proportion of processes are in the $r=0$ regime [6]. This would
be the case if the initial explosion of the big bang singularity was a process
of type $r=0$. All subsequent processes would then typically be of type $r=0$
with the possible exception of exceedingly rare, highly singular processes
producing a type $r=1$ event. However, the sheer preponderance of $r=0$
processes--including the ``master $r=0$ process,'' the cosmic arrow--implies
an improbably high entropy barrier that such rare $r=1$ processes must overcome.

\subsection{Causal Considerations}

One might also argue against time-reversed processes by invoking a standard
formulation of the causal relation between events: causes must precede their
effects in \textit{temporal order}. However, the more general form of the
causal relation is that causes must precede their effects in \textit{logical
order}, leaving open the possibility for backwards-in-time causation. For the
preparation-registration arrow, such causal considerations present problems
for $r=1$ type processes. The preparation of states $\phi$ is required before
observables $\psi$ can be measured because observables logically presuppose
states [4, 8]. The $r=1$ regime appears to contradict this causality
requirement in that observables $\psi$ are ``prepared'' before states $\phi$
can be ``measured.'' This is to say, that $R$ interchanges the roles of states
and observables. If observables are logically dependent on states, then one
might argue that there must be some kind of (strange) state in the $r=1$
regime for $t>0$ unaccounted in Table 2, but the production of such states
presents insurmountable difficulties (\S5.1).

For the more general case of the excitation/de-excitation arrow, causal
considerations do not necessarily rule out time-reversed states. For the $r=0$
regime, excitations $\phi_{+}$ lead to de-excitations $\phi_{-}$ (e.g. by
emitting some decay product leading to de-excitation). In contrast for the
$r=1$ regime, the transformation rules indicate that de-excitations $\phi_{-}$
lead to excitations $\phi_{+}$, as again the roles of the vector spaces become
interchanged. That suggests the identification of (5a) as a de-excitation into
the past is not unique. If we keep this latter identification of decay into
the past, there is nothing more to be said, as there is neither a temporal nor
a logical relationship specifying the order of excitation and de-excitation.
There are only the improbability considerations described above.

Even if we modify the identification as the $r=1$ regime suggests, this does
not immediately lead to an argument ruling out $r=1$ processes because again
the logical form of the causal relation does not foreclose the possibility
that de-excited states may become re-excited in a time reversed fashion. There
are two cases. First, the spontaneous excitation could be self-caused, but
this violates the causal relation in that all effects must have a cause. The
only possibility in this case is an uncaused event, sheer chance. Second, some
process leads to the spontaneous excitation of the de-excited state into the
past. The Unruh effect, where some kinds of ground states can be spontaneously
excited even when moving through vacuum, and pair production are possible
mechanisms in quantum field theory, but these effects are not immediately
applicable if we restrict ourselves to standard quantum mechanics.

\subsection{``Weirdness'' Considerations}

However, it does appear that the $r=1$ regime presents an interpretive
difficulty. Under the registration-preparation arrow, observables are now
represented by growing eigenvectors while states are represented by decaying
eigenvectors. Under the excitation/de-excitation arrow, if one follows what
the transformation rules suggest, de-excitations are represented by growing
eigenvectors while excitations are represented by decaying eigenvectors. These
associations are clearly not as natural as those in the $r=0$ regime, perhaps
suggesting some as yet undiscovered problems with the fourth representation of
Table 1.

\section{Discussion}

It appears that the time-reversal invariance of the dynamics in conventional
quantum mechanics is due to the underlying symmetries of the HS in which it is
formulated. This time-reversal symmetry is missing from the RHS generalization
for the case of resonance phenomena. Nevertheless, it may be possible to
restore some form of time-reversal symmetry in RHS quantum mechanics via the
extended spacetime symmetry group. For the registration/preparation arrow,
while the formalism allows states (and observables) to be distinguished from
their time-reversed counterparts, such counterparts are not physically
possible. For the more general excitation/de-excitation arrow, time-reversed
counterparts may also be distinguished, but appear to be only highly improbable.

However, there is something weird about the fourth extended spacetime
representation and the rareness of $r=1$ processes may be related to this
weirdness. If there turns out to be a serious problem with this representation
(e.g. a problematic unexamined assumption) such that it must be discarded,
then time-reversed states would disappear from RHS quantum mechanics in the
context of resonance phenomena as unphysical, leaving a purely time-asymmetric theory.

\section{References}

1. H. Atmanspacher and R. Bishop, A. Amann, Extrinsic and Intrinsic
Irreversibility in Probabilistic Dynamical Laws, in A. Khrennikov (ed.)
\textit{Quantum Probability and White Noise Analysis Volume XIII,} World
Scientific, Singapore (2002), 50-70.

2. I. Antoniou and I. Prigogine, \textit{Physica A} \textbf{192},\textbf{
}(1993) 443-64.

3. R. Bishop, \textit{Int. J. Theor. Phys.} (forthcoming).

4. A. Bohm, Rigged Hilbert Space and Mathematical Description of Physical
Systems, in W. Brittin, et al. (eds.) \textit{Lectures in Theoretical Physics
Vol IX A: Mathematical Methods of Theoretical Physics}, Gordon and Breach
Science Publishers, New York (1967) 255-317.

5. A. Bohm, \textit{The Rigged Hilbert Space and Quantum Mechanics}, Springer,
Berlin (1978).

6. A. Bohm, \textit{Physical Review A} \textbf{51,} (1995) 1758-69.

7. A. Bohm and M. Gadella, \textit{Dirac Kets, Gamow Vectors, and Gel'fand
Triplets, Lecture Notes in Physics, vol. 348}, Springer, Berlin (1989).

8. A. Bohm, S. Maxson, M. Loewe and M. Gadella, \textit{Physica A}
\textbf{236}, (1997) 485-549.

9. A. Bohm and S. Wickramasekara, \textit{Foundations of Physics} \textbf{27},
(1997) 969-93.

10. J. Cari\~{n}ena and M. Santander, \textit{J. Math. Phys.} \textbf{22},
(1981) 1548.

11. I. Gel'fand and G. Shilov, \textit{Generalized Functions Volume 3: Theory
of Differential Equations}, M. Mayer (tr.), Academic Press, New York (1967).

12. I. Gel'fand and N. Vilenkin, \textit{Generalized Functions Volume 4:
Applications of Harmonic Analysis}, A. Feinstein (tr.), Academic Press, New
York (1964).

13. C. George, \textit{Bulletin de la Classe des Sciences Academie Royale des
Sciences, des Lettres et des Beaux-Arts de Belgique} \textbf{56}, (1971), 505

14. P. Lax and R. Phillips, \textit{Scattering Theory},\textit{ }Academic
Press,\textit{ }New York (1967).

15. T. Lee, \textit{Particle Physics and Introduction to Field Theory},
Harwood Academic Publishers, New York (1981).

16. G. Ludwig, \textit{Foundations of Quantum Mechanics, Vol. I}, Springer,
Berlin (1983).

17. G. Ludwig, \textit{Foundations of Quantum Mechanics, Vol. II}, Springer,
Berlin (1985).

18. J. von Neumann, \textit{Mathematical Foundations of Quantum Mechanics},
Princeton University Press, Princeton (1955/1932).

19. J. Wheeler and W. Zurek, \textit{Quantum Theory and Measurement},
Princeton University Press, Princeton (1988).

20. E. Wigner, Unitary Representations of the Inhomogeneous Lorentz Group
Including Reflections, in \textit{Group Theoretical Concepts and Methods in
Elementary Particle Physics}, F. G\"{u}rsey (ed.), Gordon and Breach, Science
Publishers, New York: (1964).

21. H. Zeh, \textit{The Physical Basis of the Direction of Time}, Third
Edition, Springer, Berlin (1999).
\end{document}